
\chapter{Introduction}
\normalspace

     Great efforts to detect Massive Compact Objects (Machos)  by observing
gravitationally lensed stars located in the Galactic bulge are being
carried out and more than 50 candidate lensing events have already been
reported  by the OGLE (Udalski et al.\ 1994) and MACHO (Alcock et al.\ 1994;
D.\ Bennet, 1994 private communication).
Their initial estimates of the optical depth $\tau \sim 3-4 \times 10^{-6}$
substantially exceed the optical depth predicted for a standard (no-missing
mass)
disk $(\tau \sim 6 \times 10^{-7})$ (Paczy\'nski 1991; Griest et al.\ 1991)
together with a barred bulge
$(\tau \sim 1.3 \times 10^{-6})$ (Paczy\'nski et al.\ 1994; Han \& Gould 1994;
Zhao et al.\ 1994).
If the bulge is not barred the predicted optical depth is much lower.

     Recent observations provide strong evidence that the bulge is
bar-shaped with its near end in the first Galactic quadrant
(Binney et al. 1991; Blitz \& Spergel 1991; Blitz 1993; Stanek et al. 1994)
although there is no consensus over the axis ratios, scale-height
for each axis, or the angle at which we view the major axis.
This disagreement reflects the difficulties of extracting information
about the structure of the inner Galaxy from conventional observations:
the system is a complex mixture of ages, velocity dispersions, and
metallicities,
and the effects of large and irregular obscuration are difficult to model.

     Although both the high lensing event rate and other observations favor
a barred bulge model, we still do not have decisive evidence confirming
the bar structure in the bulge.
The high event rates could be ascribed to other possible structures
such as substellar objects in either a dark disk or a flattened dark halo.
That is, a high event rate does not necessarily imply that the bulge is barred.
Even after adopting barred bulge model, there is still a significant
unexplained excess of $\tau$.

     However, one could get much more information about the geometry of lens
distribution if the microlensing rate is measured over a large range of
Galactic longitude and latitude (Paczy\'nski et al.\ 1994).
Evans (1994) constructed a microlensing map (contours of
equal optical depths or rates of events with respect to $l$ and $b$).
However, to construct his map Evans tentatively adopted a very simplified
barred bulge model because no observationally well-constrained
data were available.
However, Dwek et al.\ (1994) have recently constructed a new model of
bulge light based on {\it Cosmic Background  Explorer (COBE)} data.
Their result shows that the bulge is bar-shaped and the bar is rotated in the
plane with its near side in the first Galactic quadrant making an angle of
$\theta \sim 20^{\circ}$ between its major axis and the line of sight to the
Galactic center.
They also found that a Gaussian (or exponential) radial density distribution
better describes the mass density distribution than does a simple power law
function.
Stanek (1994) proposed that one could constrain the disk/bulge normalization
by measuring the systematic offset in the apparent magnitude between lensed
clump giant stars and all observed clump giants.
This effect arises because stars from the far side of the bar are much more
likely to be lensed than the stars from the near side of the bar.
If the effect is measured and detected, one can confirm that Galactic bulge
stars
are important sources of microlenses as suggested by Kiraga \& Pacz\'nski
(1994).
However, one could not obtain information about the geometry of the bulge mass
distribution.

     Here we address two questions.
First, how real is the apparent discrepancy between the observed and
predicted lensing rates?
That is, could the observed excess just be due to statistical fluctuations?
A naive answer would appear to be no, since there have been $N \sim 50$ events
observed, so the Poisson error should be $\sim N^{-1/2} \sim 14 \%$.
However, we find that the naive error estimate is far too optimistic.
Second, how many events must be observed in the future in order to confirm,
from  microlensing, the bar-like nature of the bulge?
Related to this, how many events are required to measure the relative
disk/bulge
normalization?
We find that one can distinguish a barred bulge from a spherical one and
measure the relative disk normalization at $3\sigma$ level by observing $\sim
200$
and $\sim 800$ events respectively, provided that the optical
depths are measured over a wide range of positions around the Galactic center.

\chapter{Estimate of Uncertainty in Optical Depth Measurements}

     Suppose that a set of lensing events are detected in a given field in
which
$N_{\star}$ stars are observed over a time interval $T$.
The optical depth can then be estimated,
$$
\tau = {\pi \over 2N_{\star}T} \sum_{j=1}^{N} { t_{e,j} \over
\epsilon_{j} },
\eqno(2.1)
$$
where $\epsilon_{j}$ is the efficiency of the detecting the event, and $N$ is
number of events detected toward a window.
The measured Einstein ring crossing time $t_{e}$ is given by,
$$
t_{e}^{2} = { 4GMD \over v^{2} c^{2} }, \ \ \ D \equiv {D_{ol}D_{ls} \over
D_{os}},
\eqno(2.2)
$$
where $D_{ol}$, $D_{os}$, and $D_{ls}$ are the distances between the observer,
lens, and source.
We assume that the efficiency is a function only of the length of the event
duration
and bin the data by duration.
Equation (2.1) can then be written
$$
\tau (x_{k}) = {\pi \over 2N_{\star}T} \sum^{N_{bin}}_{i=1} n_{i}P_{i},\ \ \ \
\
P_{i} = { (t_{e})_{i} \over \epsilon_{i} }
\eqno(2.3)
$$
where $N = \sum_{i=1}^{N_{bin}} n_{i}(t_{e})_{i}$.
The square of the expected fractional error in $\tau $ is by definition
$$
{ \sigma_{\tau}^{2} \over \VEV{\tau }^{2} }
= { \VEV{ (\sum_{i}n_{i}P_{i})^{2} } - \VEV{ \sum_{i} n_{i}P_{i} }^{2} \over
\VEV{ \sum_{i} n_{i}P_{i} }^{2} } .
\eqno(2.4)
$$
Since the $P_{i}$ are constants
and the lensing events obey Poisson statistics,
$$
\VEV{ n_{i}n_{j} } = \VEV{ n_{i} } \VEV{ n_{j} } + \delta_{ij} \VEV{ n_{i} },
\eqno(2.5)
$$
equation (2.4) can be evaluated,
$$
{ \sigma_{\tau}^{2} \over \VEV{\tau }^{2} } =
{ \VEV{ \sum_{i} n_{i}P_{i}^{2} } \over \VEV{ \sum_{i} n_{i}P_{i} }^{2} } =
{ 1 \over N } { \VEV{P^{2}} \over \VEV{P}^{2} }.
\eqno(2.6)
$$
Given enough observations, it is possible to evaluate equation (2.6) from
the data themselves.
Here we make an analytic estimate by assuming that the efficiency is
independent
of the duration:
$$
{ \sigma_{\tau}^{2} \over \VEV{\tau }^{2} } =
{ \eta \over N},\ \ \ \ \eta = { \VEV{ t_{e}^{2} } \over \VEV{ t_{e} }^{2} }.
\eqno(2.7)
$$

     A naive estimate of optical depth error leads to
$\sigma_{\tau} / \VEV{\tau}  = 1 / \sqrt{N}$.
This would be valid for $t_{e} = {\rm constant}$ but otherwise results
in a significant underestimate.
Let us now adopt a simplified model in which masses, velocities, and positions
are  completely uncorrelated.
With the definition $(2.2)$ of $t_{e}$, equation (2.7) becomes,
$$
{ \sigma_{\tau}^{2} \over \VEV{\tau }^{2} }
= {1 \over N} { \VEV{v^{-2}} \over \VEV{v^{-1}}^{2} }
{ \VEV{D} \over \VEV{D^{1/2}}^{2} }
{ \VEV{M} \over \VEV{M^{1/2}}^{2} }.
\eqno(2.8)
$$
The (flux-weighted) velocity ratio is
$\VEV{v^{-2}} / \VEV{v^{-1}}^{2} = \pi / 2$
for a Gaussian velocity distribution.
For the bulge, $D$ may be plausibly modeled as uniform over an interval
$\left[ 0,L \right]$, where $L$ is physical dimension of the bulge.
Hence, the (flux-weighted) distance ratio is
$\VEV{D} / \VEV{D^{1/2}}^{2} = 16/15$.
The dispersion in the masses is more difficult to estimate.
However, if we adopt a power-law mass function
$dN(m) / d\ln m \propto m^{-\gamma}\ (\gamma > 1.5)$ with a lower-bound cut
off,
then $\VEV{M} / \VEV{M^{1/2}}^{2} = \{1 - [2(\gamma-1)]^{-2}\}^{-1}$.
Note that this ratio is a strong function of $\gamma$.
For $\gamma > 2$, $\VEV{M} / \VEV{M^{1/2}}^{2} < 4/3$.
On the other hand, if $\gamma \sim 1.5$ as in the solar neighborhood, then
the ratio diverges.
We write,
$$
{ \sigma_{\tau} \over \tau } = \sqrt{\eta \over N },
\ \ \ \ \eta > {8 \pi \over 15}.
\eqno(2.9)
$$
The inequality arises not only because of the mass dispersion but also because
if $\epsilon$ is a function of $t_{e}$ (as it surely is), this contributes an
additional term to $\sigma_{\tau}$.
Adopting a plausible estimate $\eta \sim 3$, the statistical error from 50
events is $\sigma_{\tau} / \tau \sim 24 \%$.
Hence, no strong conclusion can be drawn about an excess of events
from the current data.

\chapter{Models Disk and Bulge Mass Distribution}

     For the construction of the microlensing map, we adopt a barred bulge
model (Dwek et al.\ 1994) which is based on the observational data obtained
by the COBE satellite.
By comparing many possible density-distribution models to their data,
they found that the bulge is bar-shaped with axis ratios (1:0.33:0.23)
and an in-plane rotation angle $\theta \sim 20^{\circ}$.
Their density model (hereafter COBE model) is represented by
$$
\nu(r_{s}) = \nu_{0} \exp (-0.5r_{s}^{2}) \times 10^{9} \
{\rm L_{\odot}pc^{-3}},
\eqno(3.1)
$$
where $\nu_{0} = 3.66 \times 10^{7} \ {\rm L_{\odot}pc^{-3}}$,
$r_{s} = \{ [ (x'/x_{0})^{2} + (y'/y_{0})^{2} ]^{2} +
(z'/z_{0})^{4}\}^{1/4}$.
The coordinates $(x',y',z')$ are centered at the Galactic center
and the $x'$ axis represents the longest axis, and the shortest axis
($z'$ axis) is directed toward Galactic pole.
The values of the scale lengths are $x_{0} = 1.58\ {\rm kpc}$,
$y_{0} = 0.62\ {\rm kpc}$, and $z_{0} = 0.43\ {\rm kpc}$, respectively.

     Since the COBE model is expressed in terms of luminosity density,
not mass density, the mass-to-light ratio $M/L$ must be determined.
We estimate $M/L$ using the tensor viral theorem (Binney \& Tremaine 1987).
We find the velocity dispersions along the three principal axes
(Han \& Gould 1994) and find
$(\sigma_{x'},\sigma_{y'},\sigma_{z'}) =
(113.6,77.4,66.3)\ {\rm km\ s^{-1}}$,
and the projected velocity dispersions
$\sigma_{x}^{2} = \sigma_{x'}^{2} \cos^{2}\theta +
\sigma_{y'}^{2} \sin^{2} \theta$, $\sigma_{y}^{2} =
\sigma^{2}_{x'} \sin^{2} \theta + \sigma^{2}_{y'} \cos^{2} \theta$, and
$\sigma_{z} = \sigma_{z'}$.
Here, the coordinates $(x,y,z)$ have their center also at the Galactic
center, but the $x$ and $z$ axes point toward the Earth and to the Galactic
pole,
respectively.
They are related to the line-of-sight distance $d$ from the observer by:
$x = R_{0} - d \cos b \cos l$, $y = d \cos b \sin l$, and
$z = d \sin b$, and to the coordinates $(x',y',z')$ by:
$x' =   x \cos \theta + y \sin \theta$, $y' = -x \sin \theta + y \cos \theta$,
and $z' = z$, where $R_{0} = 8\ {\rm kpc}$ is the adopted distance to
the Galactic center.
To obtain the total mass of the bulge, the radial dispersion
$\sigma_{x}$ is normalized to the observed mean radial velocity dispersion
($\sim 110\ {\rm km\ s^{-1}})$, resulting in a  self-consistent total bulge
mass $M_{bulge} = 1.8\times 10^{10}\ {\rm M_{\odot}}$.
The calculated $M_{bulge}$ is somewhat higher than Dwek et al.'s
estimate of $M_{bulge} = 1.3 \times 10^{10}\ {\rm M_{\odot}}$
for main sequence stars based on a
Salpeter mass function cut off at $M=0.1\ {\rm M_{\odot}}$.
Zhao et al.\ (1994) also estimated $M_{bulge} \sim
1-2\times 10^{10}\ {\rm M_{\odot}}$.
They adopted the maximum value $M_{bulge} = 2
\times 10^{10}\ {\rm M_{\odot}}$ for their optical-depth computation
to explain the observed excess of $\tau$.
Because of their slightly higher estimate of $M_{bulge}$, their estimate of
$\tau$ through Baade's Window is somewhat larger than that of
Han \& Gould (1994).

     We analyze the influence of a barred structure in the bulge on the
microlensing
events by comparing the optical depths computed using the COBE
model with values calculated for an isothermal (spherical) bulge model.
The isothermal bulge model is represented by
$$
\rho(r) = \sigma_{bulge}^{2}/ 2\pi Gr^{2} =
36.7(\sigma_{bulge}/r)^{2}\ {\rm M_{\odot}
pc^{-3}},
\eqno(3.2)
$$
where $\sigma_{bulge}^{2}$ is the velocity dispersion of bulge stars.
There are other spherical or axisymmetric models proposed
(e.g.\ Bahcall 1986; Kent 1992).
However, the optical depth of lensing events does not strongly depend
on the density distribution models unless the model includes
the barred structure of the bulge (Han \& Gould 1994).
We therefore use the isothermal bulge model as representitive of all
axisymmetric bulge models.

     To compute the contribution to the optical depth by disk Machos
we adopt the Bahcall (1986) disk model which is represented by
$$
n(R,z) = n(0,0) \exp \left[ -\left({R-8000\over 3500} + {z\over 325} \right)
\right],
\eqno(3.3)
$$
where $R = (x^{2} + y^{2})^{1/2}\ {\rm pc}$ and
$n(0,0) = 0.097\ {\rm pc^{-3}}$.
The disk mass density distribution is obtained by normalizing to the mass
density of the solar neighborhood $\rho_{0} \sim 0.6\ {\rm M_{\odot}pc^{-3}}$.

\chapter{Microlensing Map}

     If bulge stars are distributed over distances between $d_{1}$ and
$d_{2}$, the optical depth $\tau$ of a bulge star lensed by foreground
bulge stars is
$$
\tau_{bulge} = {4\pi G \over c^{2}}
{\int_{d_{1}}^{d_{2}}dD_{os}n(D_{os})\int_{d_{1}}^{D_{ol}}
dD_{ol}  \rho (D_{ol}) D }
\left[ \int_{d_{1}}^{d_{2}} dD_{os}n(D_{os}) \right]^{-1}.
\eqno(4.1)
$$
Here, $D \equiv D_{ol}D_{ls} / D_{os}$,
$n(D_{os})$ is number density of source stars, and $\rho (D_{ol})$ is the mass
density of Machos along the line of sight.
We assume that $n\propto \rho$.
On the other hand, the optical depth of lensing events lensed by disk stars
can be approximated as
$$
\tau_{disk} = {4\pi G \over c^{2}} \int_{0}^{d_{2}} dD_{ol}
\rho (D_{ol})  D.
\eqno(4.2)
$$
This approximation is possible because the source bulge stars are located
within a narrow region compared to the typical values of $D_{ol}$ and $D_{ls}$
of disk lenses.
In our computation the detection factor $\beta$ (Kiraga \& Paczy\'nski 1994)
is assumed to be $1$: the increase of the volume element
and the decrease in the fraction of stars actually detected with increase of
distance cancel each other out.

     The microlensing maps with barred (Figure 1) and spherical (Figure 2)
bulge
models show significant differences.
First, the peak of $\tau$ of the barred bulge is shifted towards the
negative $l$ direction whereas the optical depth contours of the spherical
bulge are symmetric around Galactic center.
The other important differences are that the optical depths of the barred
bulge model are nearly twice as big as those of the spherical bulge model over
most of the area around Galactic center and the rate of increase of $\tau$
toward
Galactic center is much steeper for the barred bulge than for the spherical
bulge.
Even though the contribution by the disk Machos smears out some of
the distinctive differences (see the lower panels of each figure),
one still could determine the mass distribution
with observations of lensing events over a large range of Galactic latitudes
and longitudes.

\chapter{Applications of Microlensing Map}

     An important question to be answered is how many events are
required to determine the geometry of the mass distribution and relative
contributions to the event rate by disk and bulge Machos.
To answer this question we compute the covariance matrix $c_{ij}$
(e.g.\ Press et al.\ 1989) of normalizations of 3 possible components.
Let $\tau_{1}(x_{k})$, $\tau_{2}(x_{k})$, and $\tau_{3}(x_{k})$ be the
optical depths expected at the position at $x_{k} \equiv (l_{k},b_{k})$
due to standard models of a disk, barred bulge, and spherical bulge,
respectively.
And let $\tau (x_{k})$ be the total expected optical depth from 3 components.
For given normalization coefficients $a_{i}$,
the total expected optical depth is,
$$
\tau (x_{k}) = a_{1}\tau_{1} (x_{k})  + a_{2}\tau_{2} (x_{k}) +
a_{3}\tau_{3} (x_{k}).
\eqno(5.1)
$$
For example, if the events are due to a standard disk and barred bulge, then
one should find within errors $a_{1} = 1$, $a_{2} = 1$, and $a_{3} = 0$.
One seeks to minimize $\chi^{2} = \sum_{k} [\tau (x_{k}) -
\tau_{obs}(x_{k})]^{2} / \sigma_{\tau}(x_{k})^{2}$.
Then, the coefficients $a_{i}$ and their standard deviations are determined by
$$
a_{i} = \sum_{i} c_{ij} d_{i},\ \ \ \ {\mit\Delta} a_{i} = c_{ii}^{1/2},
\ \ \ d_{i} = \sum_{k} {\tau_{i}(x_{k}) \tau_{obs}(x_{k}) \over
\sigma_{\tau}(x_{k})^{2}},
\eqno(5.2)
$$
where the covariance matrix is
$$
c = b^{-1},\ \ \ \
b_{ij} = \sum_{k} {\tau_{i}(x_{k}) \tau_{j}(x_{k}) \over
\sigma_{\tau}(x_{k})^{2}}.
\eqno(5.3)
$$

     The uncertainties in determining the geometry of the bulge mass
distribution
and the relative contributions by the bulge and disk stars are computed from
equations (5.2) and (5.3) and the results are listed in Table 1.
The computations are based on 18 windows (Blanco \& Terndrup 1989,
Udalski et al.\ 1994, hereafter BT windows) available toward the Galactic
bulge (marked with open squares in each figure) and uniform coverage over
the range $-10^{\circ} < b < 10^{\circ}$, $350^{\circ} < l < 10^{\circ}$.
To see the effect of a much higher density of the disk as suggested by Gould,
Miralda-Escud\'e, \& Bahcall (1994), we also consider
a model (GMB model) that has the same functional form as Bahcall model but with
three times the normalization.
Under the assumption that the mass distribution of the disk follows the
Bahcall model, the expected uncertainties in determining the fraction of events
due to
disk lenses and the geometry of the bulge mass distribution
(barred or spherical) are ${\mit\Delta}a_{1} \sim 9.75(N/\eta)^{-1/2}$ and
${\mit\Delta}a_{i=2,3} \sim 8.0(N/\eta)^{-1/2}$, respectively.
The required numbers of events for these determinations at the $3\sigma$ level
are
therefore $850 \eta$ and $580 \eta$.
These large numbers arise because the BT windows cover
regions where the relative contributions from different structures are
fairly similar.
However, if one could cover a wide range of position around the bulge,
the uncertainties and required numbers of events would be significantly
reduced: one would need only $\sim 270\eta$ events to measure the fraction of
the
optical depth due to disk lenses, and $190 \eta$ events would be enough to
estimate
the geometry of the bulge mass distribution at the $3\sigma$ level or with
$\sim 30\%$ accuracy.

     For the determination of the bulge geometry (but not the disk
normalization)
it should be possible to make the measurement from the distribution of
event rates $\Gamma$ rather than optical depth $\tau$.
If this is true, $\eta \rightarrow 1 $ in all equations relating to the
bulge geometry (as presented in the Table 1).
The distributions of $t_{e}$ within the two bulge models are relatively
well constrained: by definition the mass spectrum is the same for the two
models.
While the velocity and distance distributions do differ, these are each
more or less fixed by the model.
Thus, while the dispersion in $t_{e}$ creates uncertainty in the optical depth,
it basically does not affect the determination of the bulge geometry.
On the other hand, one has no prior information of the mass spectrum of disk
Machos
relative to that of bulge Machos.
This implies that serious errors could result if the event rates were used
as proxies for the optical depths because the Einstein ring crossing time
$t_{e}$,
and therefore lensing rate, is a function of a Macho mass (see eq.\ (2.2)).
The factor $\eta$ therefore {\it does} degrade the measurement of the
disk/bulge
decomposition.

\chapter{Conclusion}

    In summary, naive Poisson statistics lead to a serious
underestimate of the uncertainty of optical depth ${\mit\Delta}\tau$.
An additional factor $\eta \sim 2$--3 arises from the dispersion of Einstein
ring
crossing times $t_{e}$.
The structure of the Galactic bulge and the relative
contributions to the lensing event rate by bulge and disk Machos could be
identified by observing gravitational lensing events over a wide range
of Galactic bulge.
The required numbers of events to make these determinations at the
$3\sigma$ level are $\sim 200$ and  $\sim 270\eta \sim 800$, respectively.
The barred and spherical bulges can be distinguished because of the significant
differences in their distributions of optical depth.
The main differences are

\noindent
1. The peak of the distribution of $\tau$ for the barred bulge model is shifted
toward the negative $l$ direction, and is centered around
$(l,b) = (359^{\circ},0^{\circ})$,
whereas the distribution for the spherical bulge model is concentric around
the Galactic center.

\noindent
2. The typical values of $\tau$ for the barred bulge model are significantly
(nearly twice) higher and more importantly the increase in $\tau$ toward the
Galactic
center is steeper than that of spherical bulge model in most of the
region around Galactic center.

\noindent
3. The events contributed by disk stars smear out the differences
between the two bulge models.
However, the differences still can be identified.

{\bf Acknowledgement}: We would like to thank G. Tiede, and
M. Everett for very helpful discussions.

\endpage

\ref{Alcock, C. et al. 1994, ApJ, submitted}
\ref{Bahcall, J. N. 1986, ARA\&A, 24, 577}
\ref{Binney, J., Gerhard, O. E., Stark, A. A., Bally, J., \& Uchida, K. I.
1991,
MNRAS, 252, 210}
\ref{Binney, J., \& Tremaine, D. M. 1987, Galactic Dynamics (Princeton
University Press, Princeton), 67}
\ref{Blanco, V. M., \& Terndrup, D. M. 1989, AJ, 98, 843}
\ref{Blitz, L. 1993, Back To The Galaxy (AIP Conference Proceedings, New York),
98}
\ref{Blitz, L., \& Spergel, D. N., 1991, ApJ, 379, 631}
\ref{Dwek, E., Arendt, R. G., Hauser, M. G., Kelsall, T., Lisse, C. M.,
Moseley, S. H., Silverberg, R. F., Sodroski, T. J., \& Weiland, J. 1994, ApJ,
submitted}
\ref{Gould, A., Miralda-Escud\'e, J., \& Bahcall, J. N. 1994, ApJL, 423, L105
}
\ref{Griest, K. et al.\ 1991, ApJ, 387, 181}
\ref{Han, C., \& Gould, A. 1994, ApJ, submitted}
\ref{Kent, S. M. 1992, ApJ, 387, 181}
\ref{Kiraga, M., \& Paczy\'nski, B. 1994, ApJL, 430, 101}
\ref{Paczy\'nski, B. 1991, ApJ, 419, 648}
\ref{Paczy\'nski, B., Stanek, K. J., Udalski, A., Szyma\'nski, M., Kalu\.zny,
J.,
Kubiak, M., Mateo, M., \& Krezmi\'nski, W. 1994, ApJL, submitted}
\ref{Press, W. H., Flannery, B. P., Teukolsky, S. A., \& Vetterling, W. T.
1989, Numerical Recipes (Cambridge University Press, Cambridge), 534}
\ref{Stanek, K. Z. 1994, preprint}
\ref{Udalski, A., Szyma\'nski, M., Stanek, K. Z., Kalu\.zny, J., Kubiak, M.,
Mateo, M., Krzemi\'nski, B., \& Venkat, R. 1994, Acta Astron., 44, 165}
\ref{Zhao, H., Spergel, D. N., \& Rich, R. M. 1994, ApJ Letters, submitted}

\refout
\endpage
\endpage
\bye